\documentclass[twocolumn]{aastex62}

\received{}
\revised{}
\accepted{}

\shorttitle{Discovery of a Molecular Collision Front in NGC 4567/4568}
\shortauthors{Kaneko et al.}

\begin{document}

\title{Discovery of a Molecular Collision Front \\
	in Interacting Galaxies NGC 4567/4568 with ALMA}

\correspondingauthor{Hiroyuki KANEKO}
\email{kaneko.hiroyuki@nao.ac.jp}

\author[0000-0002-2699-4862]{Hiroyuki KANEKO} 
\affil{Nobeyama Radio Observatory, 462-2, Nobeyama, Minamimaki, Minamisaku, Nagano, 384-1305, Japan} 
\affiliation{National Astronomical Observatory of Japan,  2-21-1 Osawa, Mitaka, Tokyo, 181-8588, Japan}

\author{Nario KUNO} 
\affiliation{Graduate School of Pure and Applied Sciences, University of Tsukuba, 1-1-1 Tennodai, Tsukuba, Ibaraki, 305-8577, Japan} 
\affiliation{Tomonaga Center for the History of the Universe, University of Tsukuba, 1-1-1 Tennodai, Tsukuba, Ibaraki 305-8571, Japan}

\author[0000-0001-8226-4592]{Takayuki R. SAITOH} 
\affiliation{Earth-Life Science Institute, Tokyo Institute of Technology, 2-12-1, Ookayama, Meguro, Tokyo, 152-8550, Japan}

\begin{abstract}
We present results of $^{12}$CO({\itshape J} = 1--0) imaging observations of \object{NGC 4567/4568}, a galaxy pair in a close encounter, with Atacama Large Millimetre/submillimetre Array. 
For the first time, we find clear evidence of a molecular collision front whose velocity dispersion is 16.8$\pm$1.4 km s$^{-1}$ at the overlapping region owing to high spatial and velocity resolution.
By integrating over the velocity width that corresponds to the molecular collision front, we find a long filamentary structure with a size of 1800 pc $\times$ 350 pc at the collision front.
This filamentary molecular structure spatially coincides with a dark-lane seen in {\itshape R}-band image.
We find four molecular clouds in the filament, each with a radius of 30 pc and mass of 10$^{6} M_{\odot}$; the radii matching a typical value for giant molecular clouds and the masses corresponding to those between giant molecular clouds and giant molecular associations. All four clouds are gravitationally bound.
The molecular filamentary structure and its physical conditions are similar to the structure expected via numerical simulation.
The filament could be a progenitor of super star clusters.
\end{abstract}


\section{Introduction} \label{sec:intro}
Gravitational interactions of galaxies (collisions and mergers) play an important role in the evolution of galaxies. 
A close galaxy-galaxy interaction event greatly alters the distribution and kinematics of stars and gas, resulting in the formation of elliptical galaxies \citep{Toomre77}.
During this phenomenon, significant enhancement of star formation activity is observed \citep{Bushouse86}.
It is also known that the activation of star formation follows the progression of the interaction:
while the star formation rate (SFR) in interacting galaxies during the early stage is only a few times higher than in field galaxies, the SFR becomes 10--100 times higher than field galaxies during the late stage of the interaction \citep{Kennicutt87, Teyssier10}.
In particular, ultra-luminous infrared galaxies (ULIRGs: $L_{\mathrm{IR}} \ >$ 10$^{12} \ L_{\odot}$), most of which are thought to be in the late stage of the interaction \citep{Clements96}, show bursts of star formation.
High-resolution numerical simulations revealed that multiple nuclei in ULIRGs can not only be made by multiple major mergers, but also a single merger, and produce massive and compact star clusters \citep{Matsui12}.
In spite of these findings from observations and simulations, the detailed mechanisms of active star formation in interacting galaxies are still unclear, for example, how off-centre starbursts occur.

Since molecular gas fuels star formation, investigating how the interaction affects molecular gas properties is an important step for understanding the mechanism of the intense star formation in interacting galaxies.
Previous molecular gas observations have mainly been focused on the interacting galaxies with starbursts \citep[e.g.,][]{Whitmore14,Saito15}.
Although these observations can help us investigate the effects of starbursts on galaxies, how star formation activity is enhanced remains unexplained.
In order to inspect how the interaction influences molecular properties, molecular gas observation of interacting galaxies at the early stage is crucial because they should maintain conditions before or at the beginning of active star formation, taking into account the timescale of star formation.

The \object{NGC 4567/4568} pair is one of the ideal interacting galaxies for this purpose because of their alignment, vicinity (16 Mpc), and weakly enhanced star formation activity in their overlapping regions.
Although stellar morphology is undisturbed, the NGC~4567/4568 pair is thought to be in the early stage from their asymmetric \ion{H}{1}, $^{12}$CO({\itshape J} = 1--0) (hereafter CO), and $^{12}$CO({\itshape J} = 2--1) morphologies and smooth velocity field \citep{Iono05,Kaneko13,Nehlig16}.
\ion{H}{1} has a peak in their overlapping region and CO distribution is also distorted towards the overlapping region.
A fraction of the molecular gas mass to the total gas mass is higher in this pair than that in field galaxies, implying the interaction already affects the molecular gas properties in this pair even during the early stage of the interaction \citep{Kaneko17, Nehlig16}.
Furthermore, there is not a starburst in the overlapping region, but there are some large star-forming regions traced by H$\alpha$ \citep{KKY01} and MIPS 24 $\mu$m \citep{Smith07}.
These facts indicate that the active star formation induced by the interaction has just commenced in this galaxy pair.

\section{Observations} \label{sec:obs}
CO observations towards the NGC~4567/4568 galaxy pair were carried out as an ALMA Cycle 1 program (2012.1.00759.S).
The 12-m array observations were performed using 36 antennas with C32-4 configuration.
Total observed time on the source was 25.4 minutes, and 39 fields of view were required to cover \replaced{the}{a} whole region of the galaxy pair.
Bandpass and phase were calibrated with J1229+0203 (the observed flux: 6.3 Jy) and J1239+0730 (the observed flux: 0.87 Jy), respectively.

Observations with Atacama Compact Array (ACA), which consists of ten 7-m antennas and two 12-m antennas for Cycle 1 observation, were also requested to image large-scale structures.
The total on-source time for the ACA 7-m array observations as 20.9 minutes. 
The bandpass calibrator and the phase calibrator for the ACA 7-m array were J1058+0133 with the observed flux of 2.7 Jy and J1229+0203 with the observed flux of 5.9 Jy, respectively.
While the ACA 7-m array observations were successfully completed, the ACA 12-m (Total Power) single-dish array observation was not performed for this program.
Interferometric data without single-dish data has no sensitivity on a structure larger than $\sim$0.6 $\lambda/L_{\textrm{min}}$ where $L_{\textrm{min}}$ is the shortest projected baseline.
For this reason, our ALMA data cannot detect emission from extended structures larger than $\sim$20$^{\prime\prime}$ (corresponding to 1.55 kpc at 16 Mpc) since $L_{\textrm{min}}$ is $\sim$15-m for our observations.
This scale is almost comparable to our previous single-dish data obtained with the Nobeyama 45-m telescope \citep{Kaneko13}.

Data reduction including calibration and imaging was made with the Common Astronomy Software Applications package (CASA) ver.4.2.1.
The 12-m array data and the ACA 7-m array data were combined together, continuum was subtracted, and corrected for the primary beam attenuation.
The imaging was done interactively using CLEAN task with Briggs weighting (robustness parameter of 0.5).
The final data were imaged on a 540 $\times$ 648 pixel with a grid size of 0$\farcs$5 pixels. 
The combined cube has an angular resolution of 2$\farcs$0 $\times$ 2$\farcs$0 (equivalent to 155 pc $\times$ 155 pc at 16 Mpc) and a velocity resolution of 5 km s$^{-1}$, which is the highest spacial and velocity resolution ever obtained in molecular lines for this galaxy pair.
Resultant rms noise is 8.5 mJy beam$^{-1}$.

\section{Results} \label{sec:result}
\begin{figure}
	\begin{center}
		\plotone{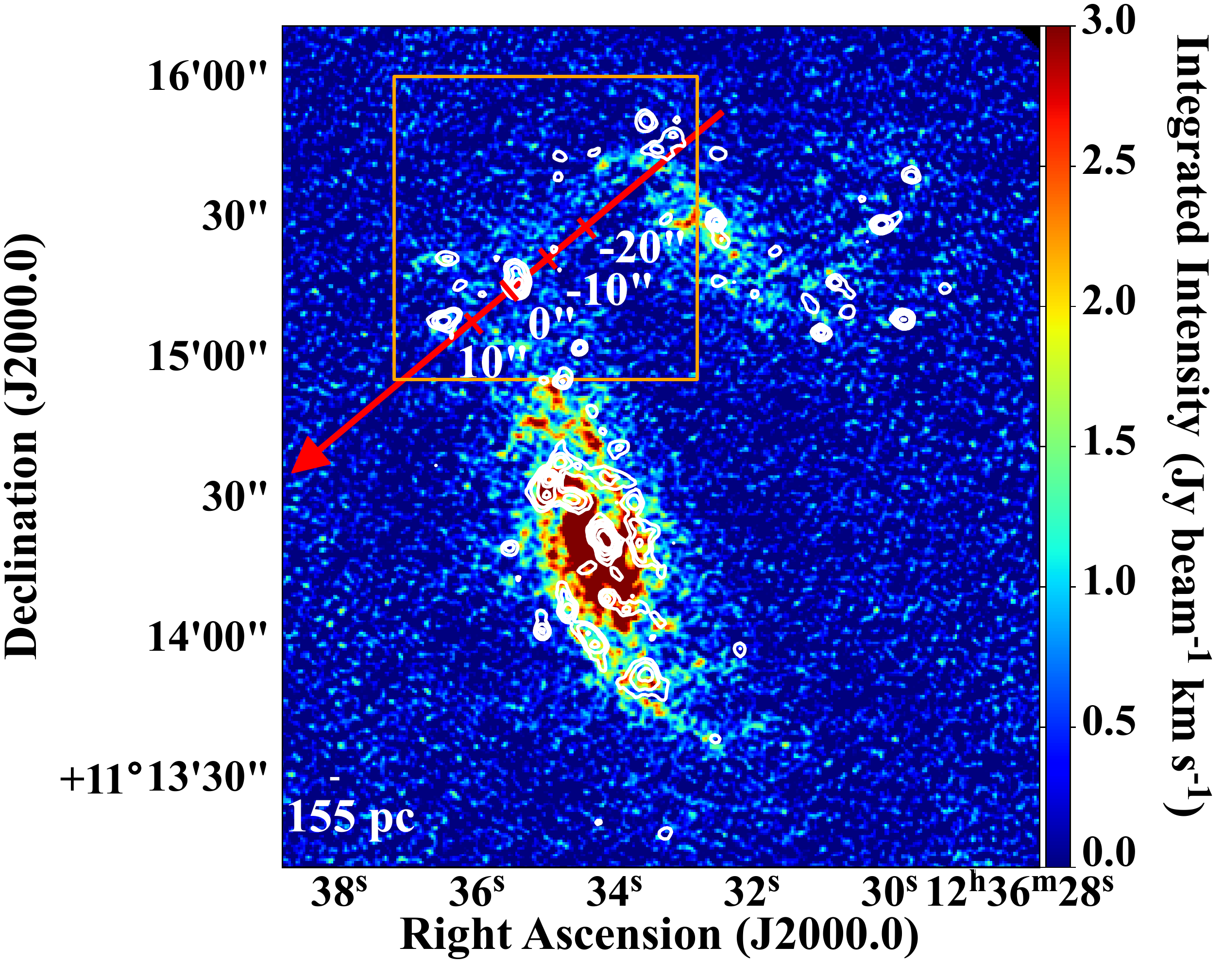}
		\caption{A CO integrated intensity map of \object{NGC 4567/4568} with H$\alpha$ contours.
			The contour levels are 2.5, 10.0, 22.5, 40.0, 62.5, and 90.0 $\times  10^{-17}$ erg s$^{-1}$ cm$^{-2}$ pixel$^{-1}$, pixel =0$\farcs$67.
			The red arrow indicates the line along which we drew a position-velocity diagram.
			Ticks along the red arrow correspond to the offset in a position-velocity diagram (Figure \ref{pvd}).
			The orange box corresponds to the region magnified in Figure \ref{filament}.}
	\end{center}
	\label{integ}
\end{figure}

A CO integrated intensity map is shown in Figure \ref{integ}.
Each galaxy reveals molecular spiral arms and strong emission\added{s} at their galactic centre.
No strong emission is seen in the overlapping region, although the single-dish observations detect significant CO emission there.
This implies the presence of extended structures.
The total CO flux from the whole system obtained with the ALMA Cycle 1 observation is (1.29 $\pm$ 0.20) $\times$ 10$^{3}$ Jy km s$^{-1}$.
Since the data observed with the ALMA Cycle 1 program has less sensitivity for the extended structure as mentioned in section \ref{sec:obs}, we estimated the missing flux by comparing to the data obtained with a single-dish telescope.
CO imaging with the single-dish was performed by our previous observations using the Nobeyama 45-m telescope\footnote{Since the velocity resolution of the single-dish data is 4 times worse than that of the ALMA data (i.e., 20 km s$^{-1}$), we cannot combine the single-dish data with ALMA data to recover the total flux.}.
The CO flux of the NGC~4567/4568 system obtained with the Nobeyama 45-m telescope is (2.38 $\pm$ 0.37) $\times$ 10$^{3}$ Jy km s$^{-1}$, 
meaning that the missing flux for the ALMA Cycle 1 observation reaches 46 $\pm$ 11 \% of the total flux.

\begin{figure}
	\begin{center}
		\plotone{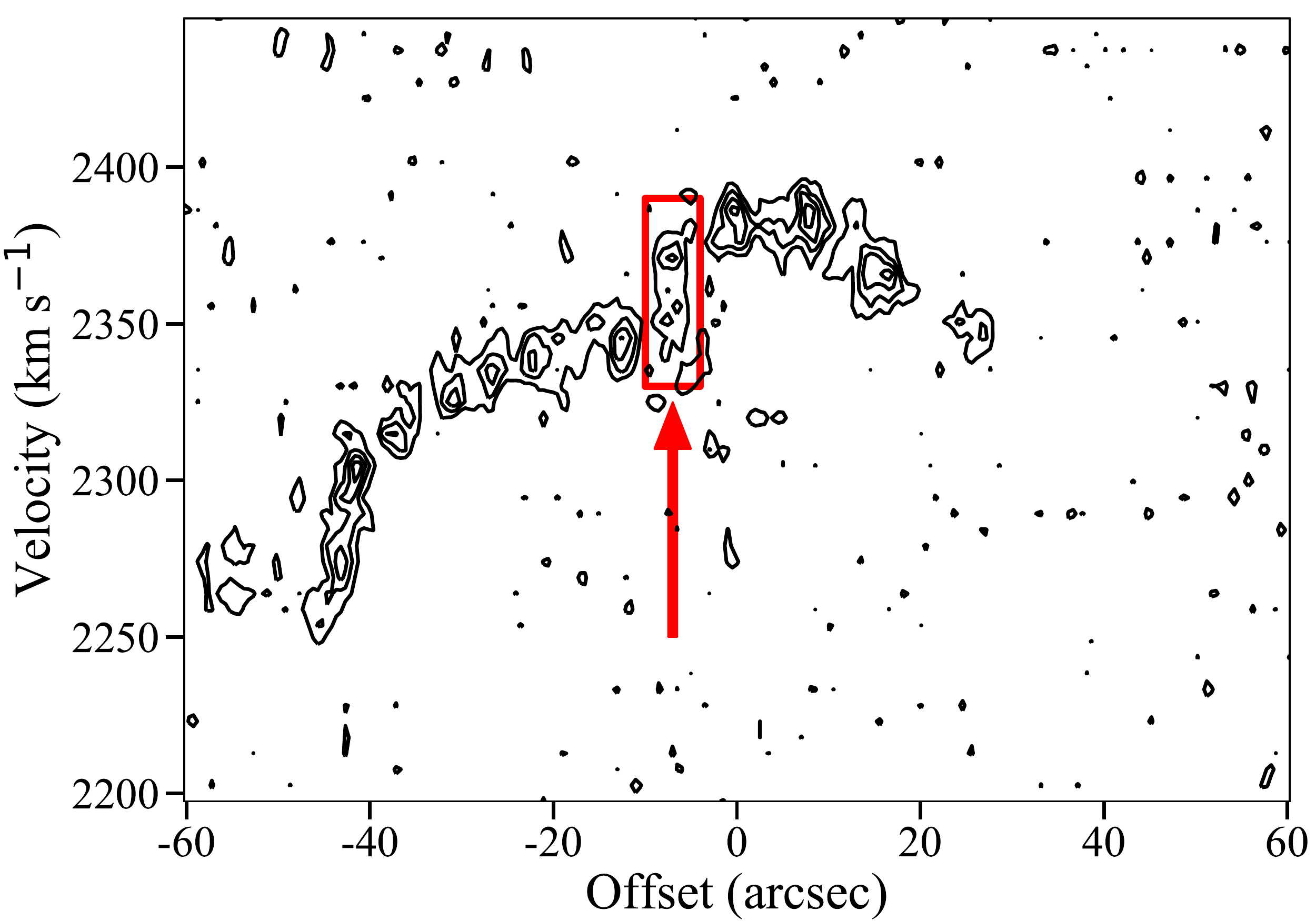}
		\caption{The position-velocity diagram along the two star-forming areas in the overlapping region. 
			The contour levels are 8.5 $\times$ 1, 2, 3, and 4 mJy beam$^{-1}$.
			Red box indicates the molecular collision front.}
	\end{center}
	\label{pvd}
\end{figure}

We made a position-velocity diagram (PVD) along with the red arrow in Figure \ref{integ} that crosses the two large star-forming areas in the overlapping region.
The line is centred on (R.A., Dec.) = (12$^{\mathrm{h}}$36$^{\mathrm{m}}$35$\fs$6, 11$^{\circ}$15$^{\prime}$15$^{\prime\prime}$) with a position angle of 50 degrees.
If two galaxies are ``apparently'' in contact with the celestial sphere and not interacting physically, the PVD should have a gap in the velocity axis.
Figure \ref{pvd} illustrates, however, molecular gas in NGC 4567 and NGC 4568 smoothly connects with velocity width of approximately 50 km s$^{-1}$ (at an offset of -5$^{\prime\prime}$ -- -10$^{\prime\prime}$).
This molecular gas having large velocity width is direct evidence of a molecular collision front, which has never been observed.

\begin{figure*}
	\begin{center}
		\plotone{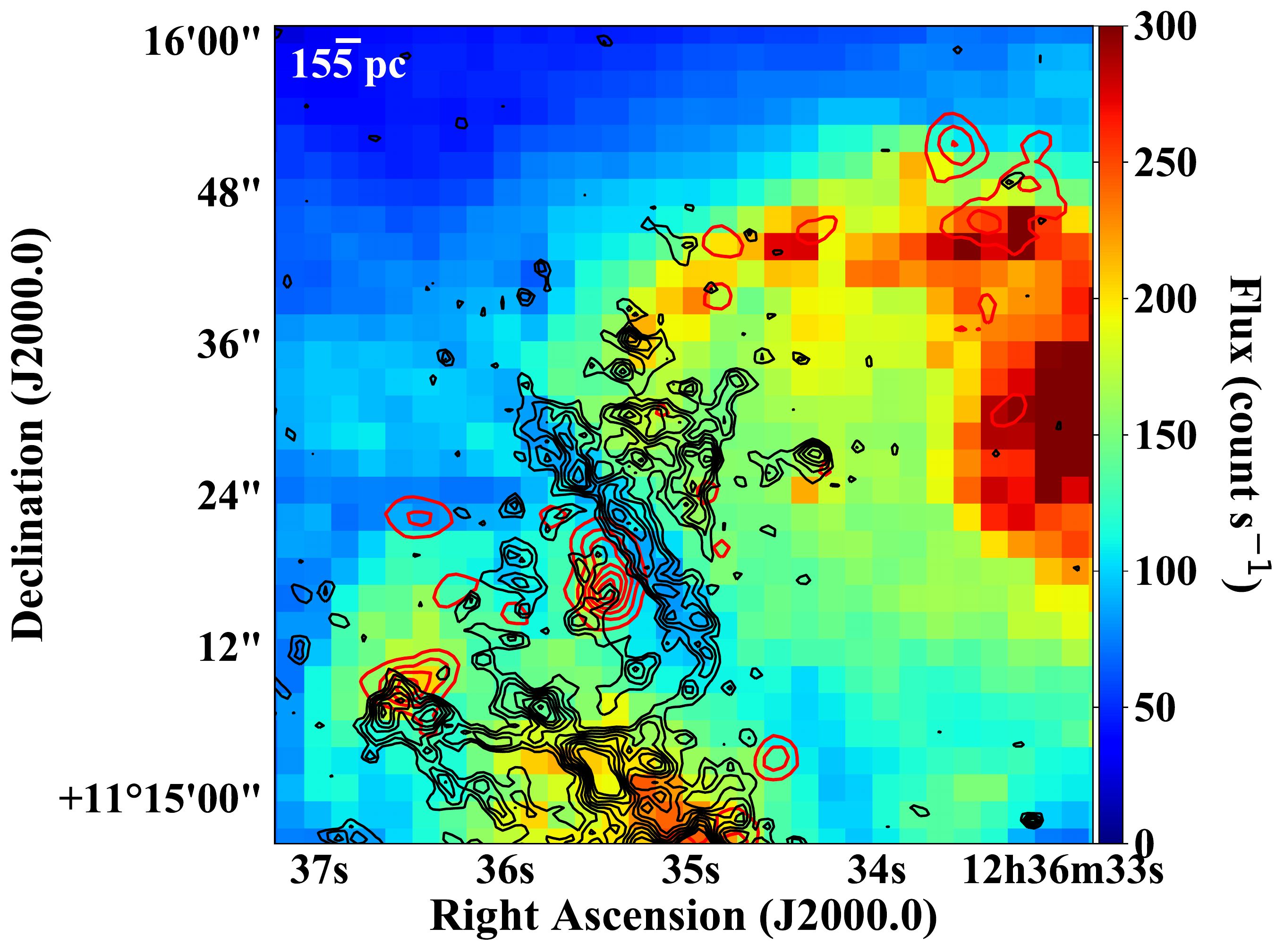}
		\caption{The close-up around the overlapping region of NGC 4567/4568. 
			Colour: SDSS $R$-band, Red Contours: H$\alpha$, Black Contours: $^{12}$CO({\itshape J} = 1--0) intensity integrated over 2330 -- 2380 km s$^{-1}$ where large velocity width is seen in Figure \ref{pvd}.}
	\end{center}
	\label{filament}
\end{figure*}

In order to reveal the structure and the distribution of the molecular collision front, 
we integrate the flux over the velocity from 2330 to 2380 km s$^{-1}$, where large velocity width is seen in Figure \ref{pvd}.
Figure \ref{filament} shows that a long filamentary structure lies adjacent to the largest star-forming domain in the overlapping region.
The filamentary structure matches a dark-lane traced by \textit{R}-band image that is located between two progenitor galaxies and does not trace spiral arms of both galaxies.
The filamentary structure contains a number of local peaks suggesting that the molecular collision front is made with an ensemble of molecular clouds.
The size of the filament is 1800 pc $\times$ 350 pc (an aspect ratio of 5), and the mass is (2.24 $\pm$ 0.07) $\times$ 10$^{7} \ M_{\odot}$ assuming the Galactic CO-H$_{2}$ conversion factor of 1.8 $\times$ 10$^{20}$ K (km s$^{-1}$)$^{-1}$ \citep{Dame01}. 
Note that the size and mass of the filament are a lower limit since about half of \added{the} CO flux is missing as we described above.

We average the spectra within the filament cloud (1800 pc $\times$ 350 pc) and fit it with the Gaussian function. Full width at half maximum (FWHM) of the filament cloud is 39.5 $\pm$ 3.2 km s$^{-1}$. Therefore the line-of-sight velocity dispersion $\sigma$ is 16.8 $\pm$ 1.4 km s$^{-1}$.
Regarding $\sigma$ as the averaged squared velocity dispersion $\left\langle\sigma^{2}\right\rangle^{1/2} $, 
we can calculate the virial mass per unit length based on \citet{FP00}:
\begin{equation}
	m_{\rm{vir}}/l = \frac{2\left\langle\sigma^{2}\right\rangle}{G},
\end{equation}
where {\itshape G} is the gravitational constant.
We have $m_{\mathrm{vir}}/l$ of (1.3 $\pm$ 0.01) $\times$ 10$^{5} \ M_{\odot} \ \mathrm{pc}^{-1}$.
Molecular gas mass per unit length, $m_{\mathrm{CO}}/l$, is 2.24 $\times$ 10$^{7}$ $M_{\odot}$/1800 pc = 1.2 $\times$ 10$^{4} \ M_{\odot} \ \mathrm{pc}^{-1}$.
Thus, the filament is currently gravitationally unbound.

\section{Discussion and Conclusion} \label{sec:discuss}
\begin{figure}
	\begin{center}
		\plotone{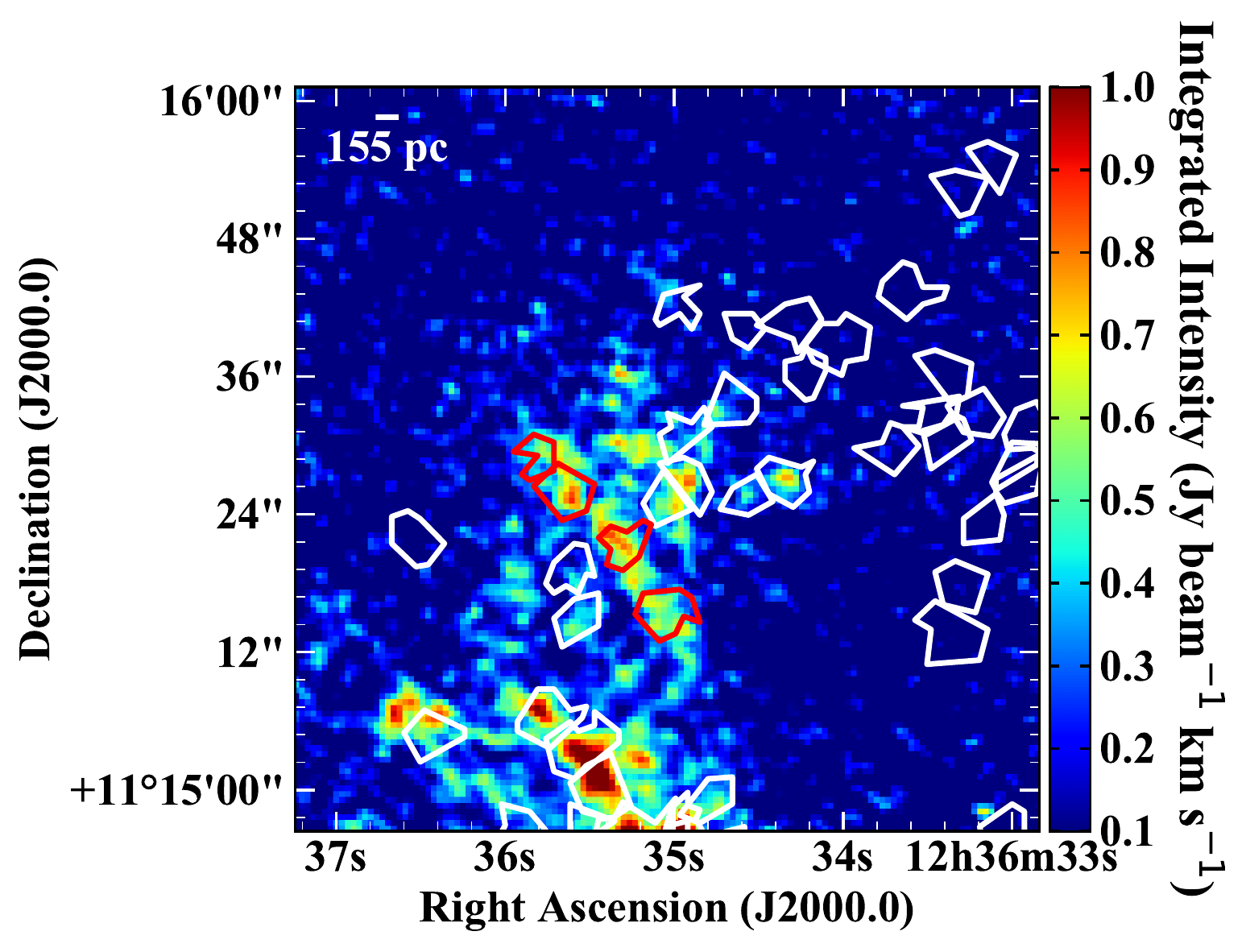}
		\caption{The close-up CO image around the overlapping region of the NGC 4567/4568 system. 
			White polygons represent non-deconvolved shapes of molecular clouds in NGC 4567/4568, while polygons enclosed by red lines show the molecular clouds in the filament.}
	\end{center}
	\label{cloud_location}
\end{figure}

In order to figure out the physical conditions of the molecular clouds in the filament, we identify the molecular clouds using CLUMPFIND software \citep{Williams95}.
With this spatial and velocity resolution data, a total of 215 clouds are resolved from the cube data, and four of them are found in the filamentary structure.
The locations and physical properties of the four clouds embedded in the filament are illustrated in Figure \ref{cloud_location} and Table \ref{clouds}, respectively.
The four clouds have a deconvolved radius of $\sim$30 pc and luminosity mass of an order of 10$^{6}\ M_{\odot}$.
The radius is comparable to the typical values for giant molecular clouds (GMCs: $\sim$10$^{5} \ M_{\odot}$, 10--100 pc) \citep{SSS85,Koda09}, while the mass is between that of GMCs and  giant molecular associations (GMAs: $\sim$10$^{7} \ M_{\odot}$, $>$100 pc) in nearby galaxies \citep{DM13,Tosaki07}.
We also derive the average deconvolved radius and luminosity mass for 211 molecular clouds located outside of the filament and find an average radius of $31.2\pm29.1$ pc and mass of $(9.57\pm0.03)\times10^{6} \ M_{\odot}$, respectively.
The molecular clouds in NGC~4567/4568 have higher molecular gas mass for their size \citep{Bolatto08}, and the four molecular clouds inside the filament have the same deconvolved radius and slightly smaller luminosity mass compared with other clouds in NGC 4567/4568.

We calculate the virial mass $M_{\mathrm{vir}}$ assuming that the clouds are a spherical shape with truncated $\rho \propto r^{-1}$ density profiles \citep{MacLaren88} and virial parameter $\alpha_{\mathrm{vir}}$ \citep{BM92} for the clouds:
\begin{equation}
	M_{\mathrm{vir}}=1040 \sigma^{2}R,
\end{equation}
\begin{equation}
	\alpha_{\mathrm{vir}}=\frac{5\sigma^{2}R}{GM},
\end{equation}
where {\itshape R} and {\itshape M} are the deconvolved radius and luminosity mass of the molecular cloud, respectively. 
The result shows that averaged $\alpha_{\mathrm{vir}}$ for all molecular clouds in NGC~4567/4568 is 0.27, which is significantly smaller than that of GMCs in nearby galaxies \citep{Rosolowsky07, Wong11}, and almost all clouds (206/215) are gravitationally bound ($\alpha_{\mathrm{vir}}<1$), including the four gas clouds in the filament.
The low virial parameter is due to large luminosity mass for the cloud size.
The results suggest that a galaxy collision contributes to the formation of heavier gas clouds.

We derive a free-fall time for the gas clouds in the filament by 
\begin{equation}
	t_{\mathrm{ff}} = \left(\frac{R^{3}}{GM}\right)^{1/2}.
\end{equation}
We find all four molecular clouds embedded \replaced{to}{in} the filament have the free-fall time of 10$^{6}$ yr, which is much shorter than the timescale for a galaxy merger event.
Hence, they will create an extended star-forming region.

\begin{deluxetable*}{ccccccccc}
	\tablecaption{Physical Properties of Molecular Clouds Embedded \replaced{to}{in} the Filament\label{clouds}}
	\tabletypesize{\footnotesize}
	\tablehead{
	Right Ascension & Declination & Radius\tablenotemark{$\dagger$} & Systemic Velocity & Velocity Dispersion\tablenotemark{$\dagger$} & Luminosity Mass & Virial Mass & Virial Parameter & Free-Fall Time\\
	(J2000.0) & (J2000.0) & (pc) & (km s$^{-1}$) & (km s$^{-1}$) & (10$^{4} \ M_{\odot}$) & (10$^{4} \ M_{\odot}$) & & (10$^{6}$ yr)}
	\startdata
	12$^{\mathrm{h}}$36$^{\mathrm{m}}$35$\fs$06 & 11$\degr$15$\arcmin$15$\farcs$49  & 38.6 & 2365 & 3.9 & 320 &  62.1 & 0.22 &2.3 \\
	12$^{\mathrm{h}}$36$^{\mathrm{m}}$35$\fs$27 & 11$\degr$15$\arcmin$21$\farcs$75  & 31.7 & 2370 & 1.1 & 122 & 3.6 & 0.03 &2.7 \\
	12$^{\mathrm{h}}$36$^{\mathrm{m}}$35$\fs$61 & 11$\degr$15$\arcmin$26$\farcs$12  & 26.5 & 2375 & 5.7 & 407 & 88.2 & 0.24 &1.1 \\
	12$^{\mathrm{h}}$36$^{\mathrm{m}}$35$\fs$79 & 11$\degr$15$\arcmin$29$\farcs$49  & 26.4 & 2380 & 2.5 & 87 &  17.6 & 0.23 &2.5 \\
	\enddata
	\tablenotetext{\dagger}{Radius and velocity dispersion are deconvolved values.}
\end{deluxetable*}

How was the filament in the overlapping region made? 
These physical properties of the filament at the collision front obtained from our observations are consistent with the numerical simulation by \citet{Saitoh09}.
They showed that a long filamentary gas structure ({\itshape a giant filament}) emerges during the first encounter due to shock compression.
Although the mass of the filament we found is two orders of magnitude smaller than that of the numerical simulation, an aspect ratio of their long filamentary gas structure is 5, which is comparable to that of the observed molecular filament in NGC~4567/4568.
The correspondence of the molecular filamentary structure with the dark-lane implies that the filament is denser than surrounding molecular gas as expected from the simulations.
Slow shock ($\sim$50 km s$^{-1}$) traced by CH$_{3}$OH maser is observed in the other interacting galaxies, VV~114\added{,} that is thought to be mid-stage merging galaxies \citep{Saito17}.
Thus, shock can occur during the galaxy interaction.
These facts suggest that the molecular filament in NGC~4567/4568 was induced by shock.

We discuss the future of the shock-induced molecular filament in the overlapping region in NGC~4567/4568 by comparing with the simulation.
As shown in Figure \ref{cloud_location}, no star formation signature is seen in the filament.
This could be explained by two possibilities as summarised by \citet{Tanaka18}.
One is that a collision of progenitor discs inhibits star formation activity.
A collision of clouds may stabilise them due to increase turbulent pressure \citep[e.g.,][]{Johnston14}.
\citet{Dobbs14} show that velocity dispersion of a cloud becomes larger through a cloud-cloud collision event, leading to an increase of the virial parameter of the cloud.
Recent observation also shows that a higher volume density is required to form stars in a high turbulent pressure environment \citep{Rathborne14}.
These results indicate that even if dense gas is formed by cloud collision, star formation may not occur as seen in the Galactic Centre.
The other possibility is that the filament is at an early phase of collision-triggered star formation, in which conditions of star formation are just fulfilled.
\citet{Saitoh11}, who described the evolution of the giant filament in the simulations, demonstrated that the giant filament becomes fragmented into smaller clumps.
The stars that have been produced from these small clumps will merge and finally result in super star clusters.
However, a collision of discs may ionise the pre-exist GMCs and ionised gas soon returns to molecular gas if it is dense enough \citep{Komugi12}.
If ionisation happens in the filament, star formation should be delayed.
Based on this scenario, the molecular filament made by shock could be the progenitor of the off-centre starburst and super star clusters in interacting galaxies.

Finally, we discuss where star formation takes place if the filament would form stars as simulated by \citet{Saitoh11}.
The existence of gravitationally bound molecular gas clouds in the filament suggests that the filament would already be fragmented and starbursts would occur almost coincidentally in each cloud with a timescale of 10$^{6}$ yr, which is shorter than that in the simulations.
On the other hand, some part of Long-Thin Filaments in the Antennae Galaxies reported by \citet{Whitmore14} is already star-forming and the others are not.
If the Long-Thin Filaments in the Antennae Galaxies have the same origin as the filament in NGC~4567/4568, the collision of molecular discs, this fact implies that star formation in the filament might not happen simultaneously.
Since the length of the filament in NGC~4567/4568 is about a half of that of the Antennae Galaxies, a shorter time-lag of star formation within the filament in NGC~4567/4568 is expected, meaning that clouds in the filament would have same physical properties.
The likelihood of the possibilities, including whether star formation will occur in the filament, could be clarified by investigations with our dense gas tracers and higher resolution data (Kaneko et al. in prep.).

\acknowledgments
This paper makes use of the following ALMA data: ADS/JAO.ALMA\#2012.1.00759.S. ALMA is a partnership of ESO (representing its member states), NSF (USA) and NINS (Japan), together with NRC (Canada) and NSC and ASIAA (Taiwan), in cooperation with the Republic of Chile. The Joint ALMA Observatory is operated by ESO, AUI/NRAO and NAOJ.
Data analysis was in part carried out on the open use data analysis computer system at the Astronomy Data Center, ADC, of the National Astronomical Observatory of Japan.

\facility{ALMA}

\end{document}